\documentstyle[12pt,epsf]{article}
  
   \newlength{\absize}
\begin{document}
  \newcommand\beq{\begin{equation}}
  \newcommand\eeq{\end{equation}}
  \newcommand{\preprint}[1]{%
    \begin{flushright}
      \setlength{\baselineskip}{3ex} #1
    \end{flushright}}
  \renewcommand{\title}[1]{%
    \begin{center}
      \LARGE #1
    \end{center}\par}
  \renewcommand{\author}[1]{%
    \vspace{2ex}
    {\normalsize
     \begin{center}
       \setlength{\baselineskip}{3ex} #1 \par
     \end{center}}}
  \renewcommand{\thanks}[1]{\footnote{#1}}
  \renewcommand{\abstract}[1]{%
    \vspace{2ex}
    \normalsize
    \begin{center}
      \centerline{\bf Abstract}\par
      \vspace{2ex}
      \parbox{\absize}{#1\setlength{\baselineskip}{2.5ex}\par}
    \end{center}}
  \begin{titlepage}
  \preprint{preprint Roma1-1257-99\\ hep-ph/9907501}
  
\vspace{3mm}

  \title{Matching of the Shape Function}

  \vskip 0.3cm

\begin{center}
{\large Ugo Aglietti$\; {}^{a,b,}$\footnote{\sl e-mail: 
ugo.aglietti@roma1.infn.it},
Giulia Ricciardi$\; {}^{a,}$\footnote{\sl e-mail:
giulia.ricciardi@na.infn.it}.}

\vspace{3mm}

$^a${\sl
Dipartimento di Fisica, Universit\'a di Roma {\sl ``La Sapienza''}.}\\

\vspace{2mm}

$^b${\sl INFN sezione di Roma, P.le A. Moro 2, 00185 Roma, Italy \\
}

\end{center}

\vspace{4mm}

\abstract{

The  shape function
$f(k_{+})$ describes Fermi motion effects in inclusive
semi-leptonic decays
such as $
B\rightarrow X_u+e+\nu $ near 
 the end-point of the lepton spectrum.
We compute the leading one-loop corrections to the shape function $f(k_{+})$
in the effective theory
with a hard cut-off regularization. The matching constant onto full QCD
is infrared safe, i.e. the leading infrared singularity
represented by the term log$^2k_{+}$ cancels in the difference of integrals.
We compare the hard cut-off result with the 
result in dimensional regularization,
the latter containing an additional factor 
of two in the coefficient of the log$^2k_{+}$ term, and consequently requiring an oversubtraction.
}

\end{titlepage}
\vskip2truecm

\setcounter{footnote}{0}

\newpage 

\section{Introduction}

The  shape
function
$f(k_{+})$ (see Eq. (\ref{defshape}) for the definition)
is introduced to describe inclusive
semi-leptonic decays such as 
\begin{equation}
\label{vub}B\rightarrow X_u+e+\nu 
\end{equation}
in the kinematical region with a large jet energy%
$$
E_X^2\sim O(m_B^2) 
$$
and with an ``intermediate'' invariant mass
$$
M_X^2\sim O(\Lambda _{QCD}\,m_B). 
$$
This region is close to 
 the end-point of the lepton spectrum, where
the background decay $B \rightarrow X_c l \nu$
is kinematically forbidden;
therefore this
is the interesting one for the 
determination of the  Cabibbo-Kobayashi-Maskawa matrix element  
$V_{ub}$ \cite{ACCM}.

The aim of this note is to discuss the
renormalization properties of the shape
function $f(k_{+})$ 
and to give a complete description
of the matching onto full QCD.
In order to be meaningful, an effective theory must have the same 
infrared 
behaviour of the original, high energy theory.
By computing the shape function in the double 
logarithmic approximation
with a hard cut-off, we find indeed the same 
double logarithm of
$k_+$ as in QCD.
In other words, the leading IR singularity,
 $\log^2 k_+$, cancels in the matching constant (coefficient function), 
implying the factorization of infrared physics into the shape function.

We compare 
 the hard cut-off result with the 
result of dimensional regularization.
In the latter,
it appears an additional factor 
of two in the term proportional to log$^2 k_{+}$,
implying that 
the double logarithm of $k_+$ does not cancel
 in the matching constant.
This is in contrast to naive expectations.
We show, however, that 
it does exist a  non-minimal subtraction scheme
where the matching is consistent and
the leading infrared singularity cancels.
In this scheme we are forced to include terms containing
simple logarithms of  $k_+$ (see Eq. (\ref{nonmin})), which is not worrysome
since in the effective theory $k_+$
is just an index,
labelling the operators defining
the shape function\cite{mn,log}.

\section{The shape function}

The shape function is defined as \cite{generale} 
\begin{equation}
\label{defshape}f(k_{+})\equiv \langle B(v)\mid h_v^{\dagger }\delta
(k_{+}-iD_{+})h_v\mid B(v)\rangle 
\end{equation}
where $h_v$ is the field in the HQET with velocity $v$ ($v^2=1)$
representing the heavy quark, $D_+\equiv n \cdot D$
and  $n_\mu \equiv (1,0,0,-1)$ ($n^2=0)$.
We are considering
an effective field theory (EFT)
where the heavy $b$ quark is taken
in the HQET and the light quark is taken in the LEET \cite{dg}.

The argument of $f(k_{+})$ is related to the kinematics of the process by 
\beq\label{argk}
k_{+}=-\frac{Q^2}{2\,v\cdot Q}, 
\eeq
where $Q\equiv m_Bv-q$ and $q$ is the momentum of the probe, i.e. of the
virtual $W$ boson in the decay (\ref{vub}).

In the semileptonic decay (\ref{vub}), all the hadronic information is
contained in the tensor%
$$
W_{\mu \nu }(v,Q)\equiv \sum_X\langle B|J_\mu ^{\dagger }(0)|X\rangle
\langle X|J_\nu (0)|B\rangle \,\delta ^4(p_B-q-p_X), 
$$
where the current is defined as $J_\mu (x)\equiv \bar u(x)\Gamma
_\mu b(x)$\footnote{
In the Standard Model, as  usual, $\Gamma_\mu=
\gamma
_\mu (1-\gamma _5)$.}. This tensor is proportional to the absorptive part
of the forward scattering tensor%
$$
T_{\mu \nu }(v,Q)\ \equiv -i\int d^4x\,e^{-iqx}\langle B|T\left( J_\mu
^{\dagger }(x)J_\nu (0)\right) |B\rangle 
$$
according to the optical theorem 
\begin{equation}
\label{abs}W_{\mu \nu }=-\frac 1\pi {\rm Im}\,T_{\mu \nu }. 
\end{equation}

In EFT, it is well known that  $W_{\mu \nu }$
can be expressed in terms of the shape function as 
\beq\label{WMN}
W_{\mu \nu }^{EFT}=s_{\mu \nu }\,\frac 1{2v\cdot Q}\ \ f(k_{+}) 
\eeq
where%
$$
s_{\mu \nu }\equiv \frac 12\,{\rm Tr\ }\left( \Gamma _\mu ^{\dagger }\hat
Q\Gamma _\nu P_{+}\right)  
$$
is a factor containing the spin effects and $P_{+}\equiv (1+\widehat{v})/2$
is the projector on the states with velocity $v$ and positive energy.
The term $1/2 v \cdot Q$
 is the jacobian factor in the change of variable from $Q^2$ to $k_+$, 
as it stems from (\ref{argk}).

We find useful to introduce a light-cone function $F(k_{+})$  
\beq\label{LC}
F(k_{+})\equiv \langle B(v)\mid h_v^{\dagger }\frac 1{iD_{+}-k_{+}+i\eta
}h_v\mid B(v)\rangle 
\eeq
such that the shape function is proportional to 
the absorptive part of $F(k_{+})$ via a
relation analogous to (\ref{abs}) 
\begin{equation}
\label{defF}f(k_{+})=-\frac 1\pi {\rm Im\ }F(k_{+}). 
\end{equation}

By using Eq.(\ref{LC}), we can write  the tensor $T_{\mu \nu }$ in the 
EFT
in a form resembling Eq.(\ref{WMN})
\beq
T_{\mu \nu }^{EFT}=s_{\mu \nu }\;\;\frac 1{2v\cdot Q}\ F(k_{+}) .
\eeq

Let us observe that a shape function $f(k_{+})^{QCD}$
 and a light-cone function $%
F(k_{+})^{QCD}$
 can  be also defined  in full QCD by means of the relations:%
$$
T_{\mu \nu }^{QCD}\equiv \left( s_{\mu \nu }\;+\Delta s_{\mu \nu }\right)
\frac 1{2v\cdot Q}\ F(k_{+})^{QCD} 
$$
and%
$$
W_{\mu \nu }^{QCD}\equiv \left( s_{\mu \nu }\;+\Delta s_{\mu \nu }\right)
\frac 1{2v\cdot Q}\ f(k_{+})^{QCD}, 
$$
where $\Delta s_{\mu \nu }$ is defined as the part of the spin structure not
proportional to $s_{\mu \nu }$. The tensor $\Delta s_{\mu \nu }$ represents
the residual spin effects not described by the EFT. The tensor $T_{\mu \nu
}^{QCD}$ can be computed using the formula%
$$
T_{\mu \nu }^{QCD}=\frac{-i}2\,{\rm Tr}\left[ P_{+}{\cal M}_{\mu \nu
}\right]  
$$
where ${\cal M}_{\mu \nu }$ is the Feynman amplitude for the forward
scattering%
$$
b(v)+\gamma ^{*}(q)\rightarrow b(v)+\gamma ^{*}(q) 
$$
with the external spinors and the photon polarizations amputated.

\section{Matching}

\begin{figure}[b]   
\begin{center}
\epsfxsize=0.7\textwidth
\epsfysize=0.5\textheight
\leavevmode\epsffile{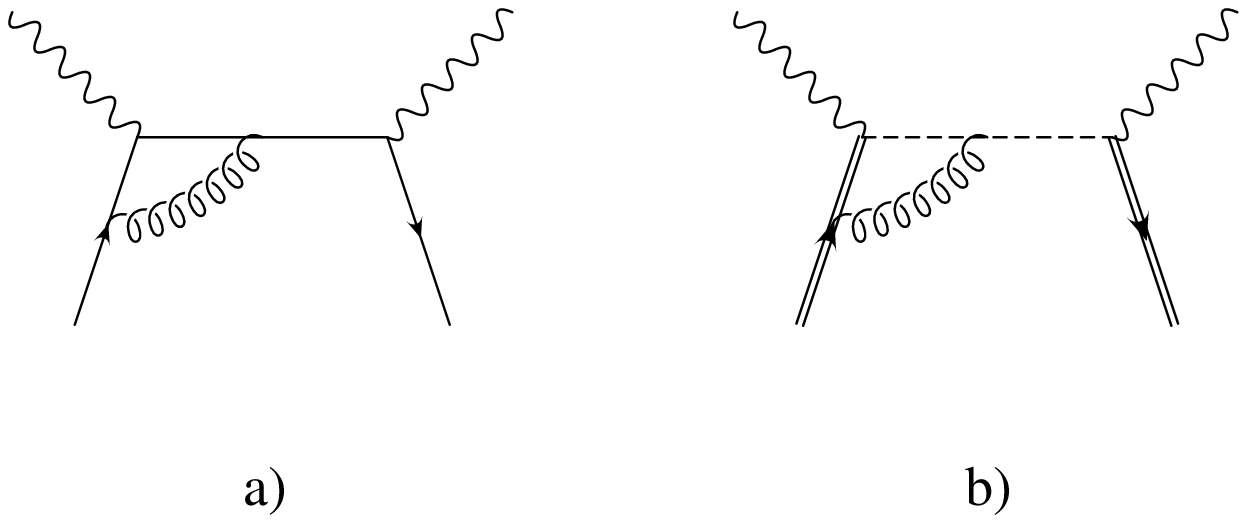}
\caption[]{\it{Vertex corrections to the light cone function
$F(k_+)$ in QCD (a) and in the EFT (b).
The external waved lines are photon lines. 
The plain 
line in a) denotes a massive or a massless fermion propagator; the gluon line 
connects the heavy and the light quarks. In b), the double line
denotes the heavy quark in the HQET, while the dashed line represents
 the light
propagator in the LEET. Both diagrams have multiplicity 2.}} 
\label{fig:dia}
\end{center}
\end{figure}
The shape function receives perturbative QCD 
corrections which determine its 
evolution through a renormalization group equation\cite{kg,mn,us}.
The starting point of the evolution
(boundary value) is determined by matching 
EFT onto full QCD.
 The matching constant (or coefficient
function) is defined through the relation 
\begin{equation}
\label{defZ}f(k_{+})^{QCD}=Z\ f(k_{+})^{EFT} 
\end{equation}
in which both the full QCD and the EFT shape functions
are computed in perturbation theory up to a prescribed accuracy
\footnote{
In perturbation theory we are actually computing the matrix elements
with external heavy quarks states;
for example, the right-hand side of Eq.(\ref{defshape}) is replaced by
$\langle b(v)\mid h_v^{\dagger }\delta
(k_{+}-iD_{+})h_v\mid b(v)\rangle$.}. It is
easier to compute the matching constant through the relation 
\begin{equation}
\label{Zeasy}F(k_{+})^{QCD}=Z\,F(k_{+})^{EFT} .
\end{equation}
By using relation (\ref{defF}) and the fact that $Z$ is real (and positive),
since it is a ratio of physical rates, one can indeed derive 
Eq.(\ref{defZ}).
At tree level:%
$$
F(k_{+})^{QCD}=F(k_{+})^{EFT}=\frac 1{-k_{+}+i\eta } 
$$
implying $Z=1$ as it should. Throughout the paper,
we will consider one-loop corrections in the
double logarithmic approximation (DLA). 
In covariant gauges, the leading
contributions come from the vertex correction diagrams (Fig.1);
we choose the Feynman gauge for simplicity. 
The computation of the light-cone function gives
in  QCD (Fig.1a))  
\begin{equation}\label{FullLight}
\label{qcdr}F(k_{+})^{QCD}=\frac 1{-k_{+}+i\eta }\left( -\frac 12\right) \
a\ \log {}^2\left( \frac{2\,m}{k_{+}-i\eta }\right) ,
\end{equation}
where $m$ is the $b$ quark mass, and $a\equiv \alpha _sC_F/\pi $
\footnote{
Balzereit et al.\cite{mn} have a similar result, but
with 
$m$ replaced by $\mu$ in Eq.(\ref{FullLight}).}.

In view of lattice-QCD applications, it is interesting to compute the
matching constant of the shape function $f(k_{+})$ in the EFT
regularized by a hard cut-off.
A regularization with a
hard cut-off $\Lambda $ (HC) on the spatial 
loop momenta is chosen 
$$
|\overrightarrow{l}|<\Lambda , 
$$
while  the loop energy $l_0$ varies
on the entire real axis
$$
-\infty <l_0<+\infty . 
$$

We could consider ordinary lattice QCD regularization as well, 
such as for example the Wilson action;
we expect similar results to hold in DLA \cite{me}. 

In the EFT we have (Fig.1b)) 
\begin{equation}
\label{hcr}F(k_{+})^{EFT}=\frac 1{-k_{+}+i\eta }\left( -\frac 12\right) \ a\
\log {}^2\left( \frac{2\,\Lambda }{k_{+}-i\eta }\right).
\end{equation}
We have taken the external states in the effective theory on-shell because
$k_+\neq 0$
 regulates the infrared divergences (soft and collinear ones)~\footnote{
For further details of the derivation see \cite{us}.}.

We explicitely see
that the 
light-cone function (or equivalently the shape function)
is ultraviolet divergent in the EFT,
the divergence being a
double logarithm of the cut-off,
while it is finite in QCD
(cfr. Eq.(\ref{FullLight})).
 Inserting these expressions into Eq.(\ref
{Zeasy}) we find%
\begin{eqnarray}\label{ZHC}\nonumber
 Z & = & 1-\frac 12a\left[ \log {}^2\left( \frac{2\,m}{k_{+}-i\eta }\right) 
-\log {}^2\left( \frac{2\,\Lambda }{k_{+}-i\eta }\right) \right]  
\\  
 & = & 1-a\log {}\left( \frac{\,m}\Lambda \right) 
\log {}\left( \frac{2\,\sqrt{m\,\Lambda }}{k_{+}-i\eta }\right). 
\end{eqnarray}
The double logarithm of $k_{+}$, i.e. the leading infrared singularity,
cancels in the difference of the integrals, implying the factorization of
infrared physics into the effective theory 
light-cone function. In particular, if
we choose 
\begin{equation}
\label{natural}\Lambda =m 
\end{equation}
as the matching scale, we get
\beq\label{Z1}
Z=1
\eeq
also at one loop in DLA.
We expect the physics to be independent from the ultraviolet
 regulator, as long as it is large enough; therefore,
we can change the cut-off from the value (\ref{natural}) to any other one
with an evolution equation in $\Lambda $.
 For example, in
DLA the evolution equation leads to
 the well-known exponentiation of the one-loop amplitude \cite
{jackiw}. Using our results: 
$$
F(k_{+})^{EFT}=\frac 1{-k_{+}+i\eta }\exp \left[ -\frac 12\ a\ \log
{}^2\left( \frac{2\,\Lambda }{k_{+}-i\eta }\right) \right] 
$$
and the QCD light-cone function is
\begin{eqnarray}\nonumber
F(k_{+})^{QCD}&=&\frac 1{-k_{+}+i\eta }\left\{ 1-\frac 12a\left[ \log
{}^2\left( \frac{2\,m}{k_{+}-i\eta }\right) -\log {}^2\left( \frac{2\,\Lambda }{k_{+}-i\eta }\right) \right] \right\} 
\\
&&\exp \left[ -\frac 12\ a\
\log {}^2\left( \frac{2\,\Lambda }{k_{+}-i\eta }\right) \right] . 
\end{eqnarray}
With the choice (\ref{natural}), we have the 
QCD light-cone function with all 
large double logarithms exponentiated 
$$
F(k_{+})^{QCD}=\frac 1{-k_{+}+i\eta }\exp \left[ -\frac 12\ a\ \log
{}^2\left( \frac{2\,m}{k_{+}-i\eta }\right) \right] . 
$$
In general, the effective theory resums the $\alpha ^n\log {}^k\,m/k_{+}$
terms in the form of renormalization-group (ultraviolet) logarithms $\alpha
^n\log {}^k\,\Lambda /k_{+}$ \cite{us}.

Up to now we have considered the standard EFT. It is
interesting to study a different effective 
field theory $(EFT^\prime)$
where the $b$ quark is 
treated in full QCD while the light quark is 
still taken in the LEET%
\footnote{In  DLA this is equivalent to replace the HQET field in 
Eq.(\ref{defshape})
with a scalar field of mass $m$.}\cite{many}. 
This effective theory contains
corrections of the form $k/m$
while it neglects corrections
of the form $k/ v\cdot Q$.
In this $EFT^\prime$ theory we 
have computed the light-cone function $
(\Lambda \gg m)$
\begin{eqnarray}
 F(k_{+})^{EFT^{\prime }}&=&\frac 1{-k_{+}+i\eta }
\left( -\frac 12\right) a\ \left[ \log {}^2
\left( \frac{2\,\Lambda }{k_{+}-i\eta }\right) -\log {}^2
\left( \frac{2\,\Lambda }m\right) \right]  
\\ \nonumber
&=&\frac 1{-k_{+}+i\eta }\left( -\frac 12\right) a\ \log {}\left( 
\frac m{k_{+}-i\eta }\right)
 \log {}\left( \frac{4\,\Lambda ^2}{m(k_{+}-i\eta )}\right) .
\end{eqnarray} 
The result is still ultraviolet divergent, as in standard EFT, but
the
divergence is reduced to a single logarithmic one.
 This result is in
contrast to naive power counting;
the addition of the quadratic term
in the heavy quark propagator, while reducing
the singularity, does not
 completely eliminate
the divergences of the effective theory.
It is also interesting to note  that the double logarithm of $%
k_{+}$ has the same coefficient as in QCD or in the ordinary EFT. 
In the $EFT^\prime$,
the
matching constant reads:%
$$
Z^\prime=
1-a\log \frac m\Lambda \log {}\left( \frac m{k_{+}-i\eta }\right)  .
$$
As in the usual EFT, $Z=1$ at $\Lambda =m$ (cfr. Eq.(\ref{Z1})).

Let us now go back to standard effective field theory (EFT),
but consider this time another regularization,
i.e.  dimensional regularization. The bare light-cone function
at fixed $\epsilon $ is given by (Fig.1b))
\begin{eqnarray}\label{DRF}
F(k_{+})^{EFT}&=&\frac 1{-k_{+}+i\eta } \left( -\frac a 2\right)  \frac
 {\Gamma(1+\epsilon)\Gamma(1+2\epsilon)\Gamma(1-2\epsilon)}{\epsilon ^2}
\left( \frac \mu {k_{+}-i\eta }\right) ^{2\epsilon } 
\\ \nonumber &=&
\frac 1{-k_{+}+i\eta }\ a\ \left[ -\frac 1{2\epsilon ^2}-\frac 1\epsilon
\log {}\left( \frac \mu {k_{+}-i\eta }\right) -\log {}^2\left( \frac \mu
{k_{+}-i\eta }\right) \right] 
\end{eqnarray}
where $\epsilon \equiv 2-D/2$, $D$ is the space-time dimension
and $\mu$ is the regularization scale \footnote{
Actually, in the
last member of Eq. (\ref{DRF}), $\mu ^2$ should be replaced by
 $\mu ^24\pi $ exp$\left[ -\gamma
_E\right] $  ($\gamma_E$ is the Euler constant),
 even though this rescaling does not affect DLA results.}. The above formula is in agreement with
Balzereit et al.\cite{mn}. The double and the simple pole are of 
ultraviolet
nature; the finite term containing the $\log {}^2k_{+}$ has an additional
factor two with respect to the HC result (Eq.(\ref{hcr})) or the QCD 
result (Eq.(\ref
{qcdr})).
 Computing the coefficient function according to Eq.(\ref{Zeasy}) we
have 
$$
Z=1+a\left[ \frac 1{2\epsilon ^2}+\frac 1\epsilon \log {}\left( \frac \mu
{k_{+}-i\eta }\right) -\frac 12\ \log {}^2\left( \frac{2\,m}{k_{+}-i\eta }%
\right) +\log {}^2\left( \frac \mu {k_{+}-i\eta }\right) \right] . 
$$
Unlike what happens with the HC regularization, the matching constant
above contains a double logarithm of $k_{+}$. 
This  term does not drop, no matter what the choice of the matching
scale is; for instance,
taking 
$$
\mu =2\,m 
$$
we have
$$
Z\left( \mu =2m\right) =1+a\left[ \frac 1{2\epsilon ^2}+\frac 1\epsilon \log
{}\left( \frac{2\,m}{k_{+}-i\eta }\right) +\frac 12\ \log {}^2\left( \frac{%
2\,m}{k_{+}-i\eta }\right) \right] . 
$$
This result 
has to be compared with the result (\ref{ZHC}) in the HC
regularization, where
the leading infrared singularity cancels  in the matching
 at any  value of $\Lambda$.

Let us trace the technical origin of the difference in the factor two
between the two regularizations. With the natural correspondence 
$$
\log \frac{2\Lambda }\mu \ \leftrightarrow \ \frac 1\epsilon 
$$
the HC result has the structure 
$$
\log {}^2\left( \frac{2\,\Lambda }{k_{+}}\right) \ \leftrightarrow \ \frac
1{\epsilon ^2}\left[ 1+\epsilon \log {}\left( \frac \mu {k_{+}}\right)
\right] ^2=\frac 1{\epsilon ^2}+\frac 2\epsilon \log {}\left( \frac \mu
{k_{+}}\right) +\log {}^2\left( \frac \mu {k_{+}}\right) ,
$$
while the DR result has the structure 
$$
\frac 1{\epsilon ^2}\left( \frac \mu {k_{+}}\right) ^{2\epsilon }=\frac
1{\epsilon ^2}\exp \left[ 2\epsilon \log \frac \mu {k_{+}}\right] =\frac
1{\epsilon ^2}+\frac 2\epsilon \log {}\left( \frac \mu {k_{+}}\right) +2\log
{}^2\left( \frac \mu {k_{+}}\right) . 
$$
The HC regularization involves a power function, while DR involves an
exponential function. Besides the first two terms (the double pole and the
simple one), these two functions do not coincide, the difference being in
fact a factor two in the finite parts.

In QCD the quantity $k_{+}$ is an (infrared) kinematical variable
(see (\ref{argk})), i.e. it
can be varied changing the external states. In the EFT the quantity $k_{+}$
is instead an index labelling the operators  entering the definition of the
shape-function (\ref{defshape}) \cite{mn,log}.
 This index is continuos and has the
dimension of a mass. It is natural to render $k_{+}$ adimensional taking the
ratio 
$$
\frac{k_{+}}m \,\, . 
$$
In the EFT, it is allowed to
change the regularization scheme according to a
ultraviolet finite replacement of the form 
\begin{equation}
\label{nonmin}\frac 1\epsilon \rightarrow \frac 1\epsilon +\log \frac{k_{+}}{%
2\,m}
\end{equation}
which renders the coefficient of the double logarithm of $k_{+}$ equal to
the QCD one:%
\begin{eqnarray}\nonumber
 \frac 1{\epsilon ^2}\left( \frac \mu {k_{+}}\right) ^{2\epsilon }
&=&\frac 1{\epsilon ^2}+\frac 2\epsilon \log \frac \mu {k_{+}}+2\log^2 
\frac \mu {k_{+}} 
\\  \label{newsc} 
&\rightarrow& \frac 1{\epsilon ^2}+\frac 2\epsilon \log 
\frac {\mu} {2m}+\log {}^2\frac {2m}{k_{+}}
+2\log \frac \mu {k_{+}}\log \frac {\mu} {2m}
\end{eqnarray} The simple pole 
in Eq.(\ref{newsc}) turns out to
have  a coefficient which is independent of $%
k_{+}$. 
The replacement (\ref{nonmin})
 is equivalent to a finite change of renormalization
prescription.
At $\mu =2\,m$, the right hand side of Eq.(\ref{newsc}) takes the
particularly simple form 
$$
\frac 1{\epsilon ^2}+\log {}^2\frac{2\,m}{k_{+}}. 
$$
The matching constant reads in the new non-minimal
dimensional scheme%
$$
Z=1+a\left[ \frac 1{2\,\epsilon ^2}+\frac 1\epsilon \log {}\left( \frac \mu
{2\,m}\right) +\ \log {}\left( \frac \mu {2\,m}\right) \log {}\left( \frac
\mu {k_{+}-i\eta }\right) \right].
$$
Taking $\mu =2m$ results in a matching constant containing only the
double (local) pole:%
$$
Z(\mu=2 m)=1+a\,\frac 1{2\,\epsilon ^2}. 
$$
The derivative with respect to log $k_{+}$ of the
last member in Eq.(\ref
{newsc}) (that controls the evolution \cite{mn, kg, us}) reads:%
$$
-2\log \frac \mu {k_{+}}. 
$$
The simple pole cancels as well as the dependence on the heavy quark mass.

\section{Conclusions}
We have shown
 that the shape function can be consistently matched onto 
QCD both with a hard cutoff and with 
dimensional regularization. In the former case the matching is
very simple and involves only the difference of the quantum corrections in
the two theories. In dimensional regularization a consistent matching
requires to go to the non-minimal scheme specified by Eq.(\ref{nonmin}).

\begin{center}
{\bf Acknowledgments}
\end{center}

It is a pleasure to thank G. Martinelli for suggestions. We acknowledge also
interesting
discussions with D. Becirevic, M. Ciuchini, E. Franco, M. Testa and L.
Trentadue.
G.R. thanks the theory group of Roma
 for its friendly hospitality
during the completion of this work. 
She also acknowledges the support of the MURST 
and of the Universit\'a di Roma ``La Sapienza''.

\end{document}